\documentclass[twocolumn,showpacs,preprintnumbers,superscriptaddress,aps,amsmath,amssymb,pre,10pt]{revtex4-1}

\usepackage{graphicx,easymat,chemarrow}
\usepackage{dcolumn}
\usepackage{bm}
\usepackage{color}
\usepackage{multirow}
\usepackage{listings,url}

\begin{document}

\preprint{PRE/DPXA}

\title{Detrended partial cross-correlation analysis of two nonstationary
  time series influenced by common external forces}

\author{Xi-Yuan Qian}
 \affiliation{School of Science, East China University of Science and
   Technology, Shanghai 200237, China}
 \affiliation{Research Center for Econophysics, East China University of
   Science and Technology, Shanghai 200237, China}

\author{Ya-Min Liu}
 \affiliation{School of Science, East China University of Science and
   Technology, Shanghai 200237, China}

\author{Zhi-Qiang Jiang}
 \affiliation{Research Center for Econophysics, East China University of
   Science and Technology, Shanghai 200237, China}
 \affiliation{School of Business, East China University of Science and
   Technology, Shanghai 200237, China}

\author{Boris Podobnik}
\affiliation{Center for Polymer Studies and Department of Physics,
  Boston University, Boston, MA 02215, USA}
\affiliation{Faculty of Civil Engineering, University of Rijeka, 51000 Rijeka,  Croatia}
\affiliation{Zagreb School of Economics and Management, 10000 Zagreb, Croatia}
\affiliation{Faculty of Economics, University of Ljubljana, 1000 Ljubljana, Slovenia}

\author{Wei-Xing Zhou}
 \email{wxzhou@ecust.edu.cn}
 \affiliation{School of Science, East China University of Science and
   Technology, Shanghai 200237, China}
 \affiliation{Research Center for Econophysics, East China University of
   Science and Technology, Shanghai 200237, China}
 \affiliation{School of Business, East China University of Science and
   Technology, Shanghai 200237, China}

\author{H. Eugene Stanley}
 \email{hes@bu.edu}
 \affiliation{Center for Polymer Studies and Department of Physics,
   Boston University, Boston, MA 02215, USA}

\date{\today}

\begin{abstract}
  When common factors strongly influence two power-law cross-correlated time series recorded in complex natural or social systems, using classic detrended cross-correlation analysis (DCCA) without considering these common factors will bias the results. We use detrended partial cross-correlation analysis (DPXA) to uncover the intrinsic power-law cross-correlations between two simultaneously recorded time series in the presence of nonstationarity after removing the effects of other time series acting as common forces. The DPXA method is a generalization of the detrended cross-correlation analysis that takes into account partial correlation analysis. We demonstrate the method by using bivariate fractional Brownian motions contaminated with a fractional Brownian motion. We find that the DPXA is able to recover the analytical cross Hurst indices, and thus the multi-scale DPXA coefficients are a viable alternative to the conventional cross-correlation coefficient. We demonstrate the advantage of the DPXA coefficients over the DCCA coefficients by analyzing contaminated bivariate fractional Brownian motions. We calculate the DPXA coefficients and use them to extract the intrinsic cross-correlation between crude oil and gold futures by taking into consideration the impact of the US dollar index. We develop the multifractal DPXA (MF-DPXA) method in order to generalize the DPXA method and investigate multifractal time series.  We analyze multifractal binomial measures masked with strong white noises and find that the MF-DPXA method quantifies the hidden multifractal nature while the MF-DCCA method fails.
\end{abstract}


\maketitle

\section{Introduction}

Complex systems with interacting constituents are ubiquitous in nature
and society. To understand the microscopic mechanisms of emerging
statistical laws of complex systems, one records and analyzes time
series of observable quantities. These time series are usually
nonstationary and possess long-range power-law
cross-correlations. Examples include the velocity, temperature, and
concentration fields of turbulent flows embedded in the same space as
joint multifractal measures
\cite{Antonia-VanAtta-1975-JFM,Meneveau-Sreenivasan-Kailasnath-Fan-1990-PRA},
topographic indices and crop yield in agronomy
\cite{Kravchenko-Bullock-Boast-2000-AJ,Zeleke-Si-2004-AJ}, temporal and
spatial seismic data \cite{Shadkhoo-Jafari-2009-EPJB}, nitrogen dioxide
and ground-level ozone
\cite{JimenezHornero-JimenezHornero-deRave-PavonDominguez-2010-EMA},
heart rate variability and brain activity in healthy humans
\cite{Lin-Sharif-2010-Chaos}, sunspot numbers and river flow
fluctuations \cite{Hajian-Movahed-2010-PA}, wind patterns and land
surface air temperatures
\cite{JimenezHornero-PavonDominguez-deRave-ArizaVillaverde-2010-AR},
traffic flows \cite{Xu-Shang-Kamae-2010-ND} and traffic signals
\cite{Zhao-Shang-Lin-Chen-2011-PA}, self-affine time series of taxi
accidents \cite{Zebende-daSilva-Filho-2011-PA}, and econophysical
variables
\cite{Lin-2008-PA,Podobnik-Horvatic-Petersen-Stanley-2009-PNAS,SiqueirJr-Stosic-Bejan-Stosic-2010-PA,Wang-Wei-Wu-2010-PA,He-Chen-2011-CSF}.

A variety of methods have been used to investigate the long-range
power-law cross-correlations between two nonstationary time series. The
earliest was joint multifractal analysis to study the cross-multifractal
nature of two joint multifractal measures through the scaling behaviors
of the joint moments
\cite{Antonia-VanAtta-1975-JFM,Meneveau-Sreenivasan-Kailasnath-Fan-1990-PRA,Schmitt-Schertzer-Lovejoy-Brunet-1996-EPL,Xu-Antonia-Rajagopalan-2000-EPL,Xu-Antonia-Rajagopalan-2007-EPL},
which is a multifractal cross-correlation analysis based on the
partition function approach (MF-X-PF)
\cite{Wang-Shang-Ge-2012-Fractals}. Over the past decade, detrended
cross-correlation analysis (DCCA) has become the most popular method of
investigating the long-range power-law cross correlations between two
nonstationary time series
\cite{Jun-Oh-Kim-2006-PRE,Podobnik-Stanley-2008-PRL,Podobnik-Horvatic-Petersen-Stanley-2009-PNAS,Horvatic-Stanley-Podobnik-2011-EPL},
and this method has numerous variants
\cite{Achard-Bassett-MeyerLindenberg-Bullmore-2008-PRE,Arianos-Carbone-2009-JSM,Wendt-Scherrer-Abry-Achard-2009,Qiu-Zheng-Chen-2010-NJP,Qiu-Chen-Zhong-Lei-2011-PA,Kristoufek-2013-EPJB,Kristoufek-2014-PA,Ying-Shang-2014-Fractals,Kristoufek-2015-PRE}. Statistical
tests can be used to measure these cross correlations
\cite{Podobnik-Grosse-Horvatic-Ilic-Ivanov-Stanley-2009-EPJB,Zebende-2011-PA,Podobnik-Jiang-Zhou-Stanley-2011-PRE}. There
is also a group of multifractal detrended fluctuation analysis (MF-DCCA)
methods of analyzing multifractal time series, e.g., MF-X-DFA
\cite{Zhou-2008-PRE}, MF-X-DMA \cite{Jiang-Zhou-2011-PRE}, and MF-HXA
\cite{Kristoufek-2011-EPL}.

The observed long-range power-law cross-correlations between two time series may not be caused by their intrinsic relationship but by a common third driving force or by common external factors \cite{Kenett-Shapira-BenJacob-2009-JPS,Shapira-Kenett-Jacob-2009-EPJB,Kenett-Tumminello-Madi-GurGershgoren-Mantegna-BenJacob-2010-PLoS1}. If the influence of the common external factors on the two time series are additive, we can use partial correlation to measure their intrinsic relationship \cite{Baba-Shibata-Sibuya-2004-ANZJS}. To extract the intrinsic long-range power-law cross-correlations between two time series affected by common driving driving forces, we previously developed and used detrended partial cross-correlation analysis (DPXA) and studied the DPXA exponents of variable cases, combining the ideas of detrended cross-correlation analysis and partial correlation \cite{Liu-2014}. In Ref.~\cite{Yuan-Fu-Zhang-Piao-Xoplaki-Luterbacher-2015-SR}, the DPXA method has been proposed independently, focussing on the DPXA coefficient.

Here we provide a general framework for the DPXA and MF-DPXA methods that is applicable to various extensions, including different detrending approaches and higher dimensions. We adopt two well-established mathematical models (bivariate fractional Brownian motions and multifractal binomial measures) in our numerical experiments, which have known analytical expressions, and demonstrate how the (MF-)DPXA methods is superior to the corresponding (MF-)DCCA methods.

\section{Detrended partial cross-correlation analysis}
\label{S1:DPXA}

\subsection{DPXA exponent}

Consider two stationary time series $\{x(t):t=1,\cdots, T\}$ and
$\{y(t):t=1, \cdots, T\}$ that depend on a sequence of time series
$\{z_i(t):t=1, 2, \cdots, T\}$ with $i=1, \cdots, n$. Each time series
is covered with $M_s=[T/s]$ non-overlapping windows of size
$s$. Consider the $v$th box $[l_v+1,l_v+s]$, where $l_v=(v-1)s$. We
calibrate the two linear regression models for ${\bm{x}}_v$ and
${\bm{y}}_v$ respectively,
\begin{equation}
  \left\{
  \begin{array}{ccc}
     {\bm{x}}_v = {\mathbf{Z}}_v{\bm{\beta}_{x,v}} + {\mathbf{r}}_{x,v}\\
     {\bm{y}}_v = {\mathbf{Z}}_v{\bm{\beta}_{y,v}} + {\mathbf{r}}_{y,v}
  \end{array}
  \right.,
  \label{Eq:xy:z:rxy:betas}
\end{equation}
where ${\mathbf{x}}_v=[x_{l_v+1},\ldots,x_{l_v+s}]^{\mathrm{T}}$,
${\mathbf{y}}_v=[y_{l_v+1},\ldots,y_{l_v+s}]^{\mathrm{T}}$,
${\mathbf{r}}_{x,v}$ and ${\mathbf{r}}_{y,v}$ are the vectors of the
error term, and
\begin{equation}
  {\bm{Z}}_v
  =\left(
  \begin{array}{ccc}
     {\bm{z}}_{v,1}^{\rm{T}}\\
     \vdots\\
     {\bm{z}}_{v,p}^{\rm{T}}\\
  \end{array}
  \right)
  =\left(
  \begin{array}{ccc}
     {\bm{z}}_{1}(l_v+1) & \cdots & {\bm{z}}_{p}(l_v+1) \\
          \vdots         & \ddots &       \cdots       \\
     {\bm{z}}_{1}(l_v+s) & \cdots & {\bm{z}}_{p}(l_v+s) \\
  \end{array}
  \right)
\end{equation}
is the matrix of the $p$ external forces in the $v$th box, where
$\bm{x}^{\rm{T}}$ is the transform of $\bm{x}$. Equation
(\ref{Eq:xy:z:rxy:betas}) gives the estimates $\bm{\hat{\beta}}_{x,v}$
and $\bm{\hat{\beta}}_{y,v}$ of the $p$-dimensional parameter vectors
$\bm{\beta}_{x,v}$ and $\bm{\beta}_{y,v}$ and the sequence of error
terms,
\begin{equation}
  \left\{
  \begin{array}{ccc}
     {\mathbf{r}}_{x,v} = {\mathbf{x}}_v-{\mathbf{Z}}_v{\bm{\hat{\beta}}_{x,v}}\\
     {\mathbf{r}}_{y,v} = {\mathbf{y}}_v-{\mathbf{Z}}_v{\bm{\hat{\beta}}_{y,v}}
  \end{array}
  \right..
\end{equation}
We obtain the disturbance profiles, i.e.,
\begin{equation}
  \left\{
  \begin{array}{ccc}
     R_{x,v}(k) = \sum_{j=1}^{k} r_x(l_v+j)\\
     R_{y,v}(k) = \sum_{j=1}^{k} r_y(l_v+j)
  \end{array}
  \right.,
\end{equation}
where $k=1,\cdots,s$.

\begin{figure*}[!ht]
  \centering
  \includegraphics[width=0.96\linewidth]{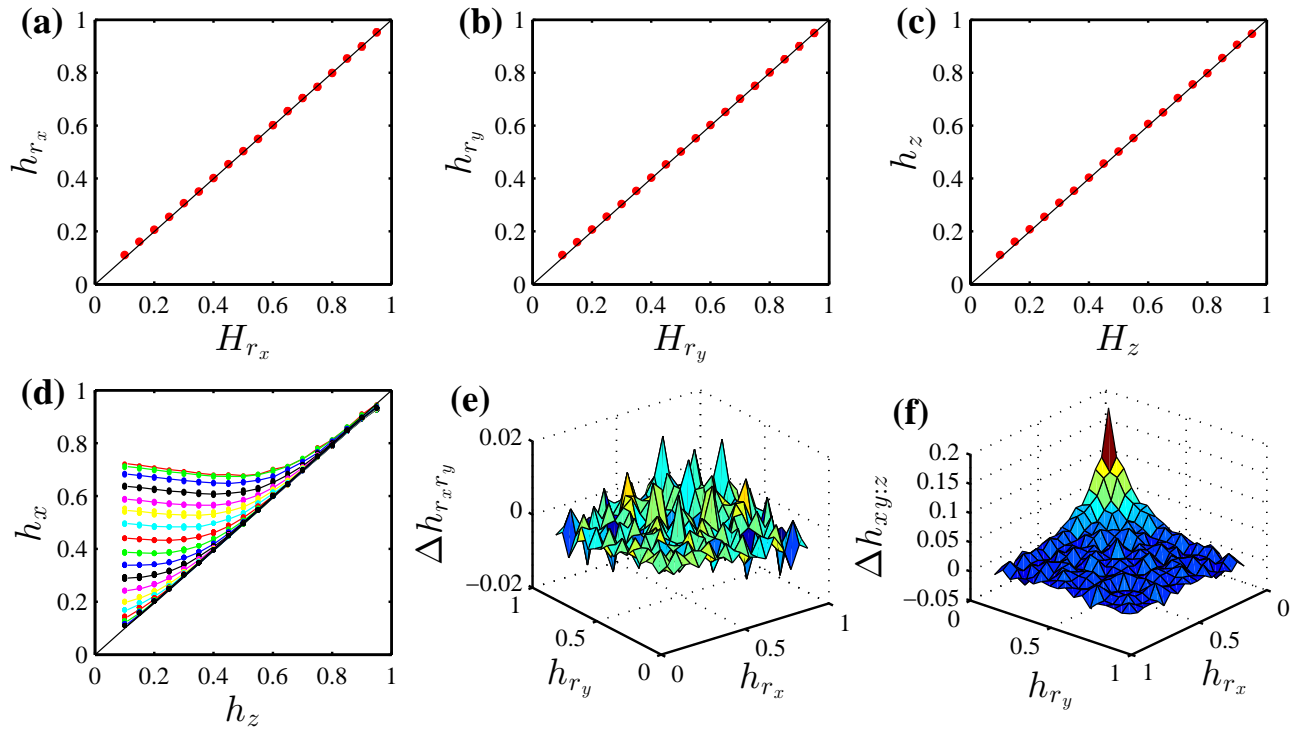}
  \caption{\label{Fig:DPXA:Dhxyz} (color online.) Detrended partial
    cross-correlation exponents. (a-b) Dependence of the average Hurst
    indices $h_{r_x}$ and $h_{r_x}$ of the two components of the
    generated bivariate fractional Brownian motions on the input Hurst
    indices $H_{r_x}$ and $H_{r_x}$. (c) Dependence of the average Hurst
    index $h_{z}$ of the generated univariate fractional Brownian
    motions on the input Hurst index $H_z$. (d) Dependence of $h_x$ on
    $H_z$ for different $h_{r_x}$ values. (e) Relative error
    $\Delta{h_{r_xr_y}} = h_{r_xr_y}-H_{r_xr_y}$. (f) Relative error
    $\Delta{h_{xy:z}}$ between the DPXA estimate
    $\langle{h_{xy:z}}\rangle_z$ and the true value $h_{r_xr_y}$ as a
    function of $h_{r_x}$ and $h_{r_y}$.}
\end{figure*}

We assume that the local trend functions of $R_{x,v}$ and $R_{x,v}$ are
$\widetilde{R}_{x,v}$ and $\widetilde{R}_{y,v}$, respectively. The
detrended partial cross-correlation in each window is then calculated,
\begin{equation}
 F_{v}^2(s) = \frac{1}{s}\sum_{k=1}^s
 \left[R_{x,v}(k)-\widetilde{R}_{x,v}(k)
   \right]\left[R_{y,v}(k)-\widetilde{R}_{y,v}(k)\right],
\end{equation}
and the second-order detrended partial cross-correlation is calculated,
\begin{equation}
 F_{xy:{\bm{z}}}(2,s) = \left[\frac{1}{m-1}\sum_{v=1}^m
   F_{v}^2(s)\right]^{1/2}.
\end{equation}
If there are intrinsic long-range power-law cross-correlations between
$x$ and $y$, we expect the scaling relation,
\begin{equation}
 F_{xy:{\bm{z}}}(2,s) \sim s^{h_{xy:{\bm{z}}}}.
 \label{Eq:Fxy:s}
\end{equation}

There are many ways of determining $\widetilde{R}_{x,v}$ and
$\widetilde{R}_{y,v}$. The local detrending functions could be
polynomials
\cite{Peng-Buldyrev-Havlin-Simons-Stanley-Goldberger-1994-PRE,Hu-Ivanov-Chen-Carpena-Stanley-2001-PRE},
moving averages
\cite{Vandewalle-Ausloos-1998-PRE,Alessio-Carbone-Castelli-Frappietro-2002-EPJB,Xu-Ivanov-Hu-Chen-Carbone-Stanley-2005-PRE,Arianos-Carbone-2007-PA},
or other possibilities \cite{Qian-Gu-Zhou-2011-PA}. To distinguish the
different detrending methods, we label the corresponding DPXA variants
as, e.g., PX-DFA and PX-DMA. When the moving average is used as the
local detrending function, the window size of the moving averages must
be the same as the covering window size $s$ \cite{Gu-Zhou-2010-PRE}.

To measure the validity of the DPXA method, we perform numerical
experiments using an additive model for $x$ and $y$, i.e.,
\begin{equation}
  \left\{
  \begin{array}{ccc}
     x(t) = \beta_{x,0} + \beta_xz(t) + r_x(t)\\
     y(t) = \beta_{y,0} + \beta_yz(t) + r_y(t)
  \end{array}
  \right.,
  \label{Eq:DPXA:Model}
\end{equation}
where $z(t)$ is a fractional Gaussian noise with Hurst index $H_z$, and
$r_x$ and $r_y$ are the incremental series of the two components of a
bivariate fractional Brownian motion (BFBMs) with Hurst indices
$H_{r_x}$ and $H_{r_y}$
\cite{Lavancier-Philippe-Surgailis-2009-SPL,Coeurjolly-Amblard-Achard-2010-EUSIPCO,Amblard-Coeurjolly-Lavancier-Philippe-2011-BSMF}.
The properties of multivariate fractional Brownian motions have been
extensively studied
\cite{Lavancier-Philippe-Surgailis-2009-SPL,Coeurjolly-Amblard-Achard-2010-EUSIPCO,Amblard-Coeurjolly-Lavancier-Philippe-2011-BSMF}. In
particular,
it has been proven that the Hurst index $H_{r_xr_y}$ of the
cross-correlation between the two components is
\cite{Lavancier-Philippe-Surgailis-2009-SPL,Coeurjolly-Amblard-Achard-2010-EUSIPCO,Amblard-Coeurjolly-Lavancier-Philippe-2011-BSMF}
\begin{equation}
 H_{r_xr_y}=(H_{r_x}+H_{r_y})/2.
 \label{Eq:Hrxry:Hrx:Hry}
\end{equation}
This property allows us to assess how the proposed method perform. We
can obtain the $h_{xy}$ of $x$ and $y$ using the DCCA method and the
$h_{xy:{\bm{z}}}$ of $r_x$ and $r_y$ using the DPXA method. Our
numerical experiments show that
$H_{r_xr_y}=h_{r_xr_y}=h_{xy:{\bm{z}}}\neq h_{xy}$. We use $H$ for
theoretical or true values and $h$ for numerical estimates.

In the simulations we set $\beta_{x,0}=2$, $\beta_x=3$, $\beta_{y,0}=2$,
and $\beta_y=3$ in the model based on Eq.~(\ref{Eq:DPXA:Model}). Three
Hurst indices $H_{r_x}$, $H_{r_y}$, and $H_z$ are input arguments and
vary from 0.1 to 0.95 at 0.05 intervals. Because $r_x$ and $r_y$ are
symmetric, we set $H_{r_x}\leq H_{r_y}$, resulting in
$\frac{(18+1)\times18}{2}\times18=3078$ triplets of $(H_{r_x}, H_{r_y},
H_z)$. The BFBMs are simulated using the method described in
Ref.~\cite{Coeurjolly-Amblard-Achard-2010-EUSIPCO,Amblard-Coeurjolly-Lavancier-Philippe-2011-BSMF},
and the FBMs are generated using a rapid wavelet-based approach
\cite{Abry-Sellan-1996-ACHA}. The length of each time series is
65536. For each $(H_{r_x}, H_{r_y}, H_z)$ triplet we conduct 100
simulations. We obtain the Hurst indices for the simulated time series
$r_x$, $r_y$, $z$, $x$, and $y$ using detrended fluctuation analysis
\cite{Peng-Buldyrev-Havlin-Simons-Stanley-Goldberger-1994-PRE,Kantelhardt-KoscielnyBunde-Rego-Havlin-Bunde-2001-PA}. The
average values $h_{r_x}$, $h_{r_y}$, $h_z$, $h_x$, and $h_y$ over 100
realizations are calculated for further analysis, which are shown in
Fig.~\ref{Fig:DPXA:Dhxyz}. A linear regression between the output and
input Hurst indices in Fig.~\ref{Fig:DPXA:Dhxyz}(a--c) yields
$\langle{h_{r_x}}\rangle=0.009+0.990H_{r_x}$,
$\langle{h_{r_y}}\rangle=0.009+0.990H_{r_y}$, and $\langle{h_z}\rangle =
0.010+0.991H_z$, suggesting that the generated FBMs have Hurst indices
equal to the input Hurst indices.  Figure~\ref{Fig:DPXA:Dhxyz}(d) shows
that when $h_{r_x}\leq h_z$, $h_x$ is close to $h_z$. When it is not,
$h_z<h_x<h_{r_x}$.

Figure~\ref{Fig:DPXA:Dhxyz}(e) shows that
$h_{r_xr_y}=(h_{r_x}+h_{r_y})/2$. Because $h_{r_x}\approx H_{r_x}$ and
$h_{r_y}\approx H_{r_y}$ [see Fig.~\ref{Fig:DPXA:Dhxyz}(a)--(b)], we verify numerically that
\begin{equation}
  h_{r_xr_y}\approx H_{r_xr_y}. 
\end{equation}
Note also that $h_{xy}\approx (h_x+h_y)/2$, and that $h_{xy:z}$ is a function of $h_{r_x}$, $h_{r_y}$ and $h_z$. A simple linear regression gives
\begin{equation}
  h_{xy:z} = 0.003+0.509h_{r_x} + 0.493h_{r_y} + 0.012h_z,
  \label{Eq:hxyz:hrx:hry:hz}
\end{equation}
which indicates that the DPXA method can be used to extract the
intrinsic cross-correlations between the two time series $x$ and $y$
when they are influenced by a common factor $z$. We calculate the
average $\langle{h_{xy:z}}\rangle_z$ over different $H_z$ and then find
the relative error
\begin{equation}
  \Delta{h_{xy:z}} = \frac{\langle{h_{xy:z}}\rangle_z-h_{r_xr_y}}{h_{r_xr_y}}.
  \label{Eq:DPXA:Dhxyz}
\end{equation}
Figure~\ref{Fig:DPXA:Dhxyz}(f) shows the results for different
combinations of $h_{r_x}$ and $h_{r_y}$. Although in most cases we see
that $\Delta{h_{xy:z}}\ll0.05$, when both $h_{r_x}$ and $h_{r_y}$
approach 0, $\Delta{h_{xy:z}}$ increases. When $h_{r_x}=h_{r_y}=0.11$,
$\Delta{h_{xy:z}}=0.192$, and when $h_{r_x}=0.11$ and $h_{r_y}=0.16$,
$\Delta{h_{xy:z}}=0.113$. For all other points of $(h_{r_x}, h_{r_y})$,
the relative errors $\Delta{h_{xy:z}}$ are less than 0.10.

\subsection{DPXA coefficient}

In a way similar to detrended cross-correlation coefficients
\cite{Zebende-2011-PA,Kristoufek-2014-PA}, we define the detrended
partial cross-correlation coefficient (or DPXA coefficient) as
\begin{equation}
  \rho_{\rm{DPXA}}(s) = \rho_{xy:z}(s) =
  \frac{F^2_{xy:{\bm{z}}}(2,s)}{F_{x:{\bm{z}}}(2,s)F_{y:{\bm{z}}}(2,s)}.
  \label{Eq:rho:DPXA}
\end{equation}
As in the DCCA coefficient
\cite{Podobnik-Jiang-Zhou-Stanley-2011-PRE,Kristoufek-2014-PA}, we also
find $-1\leq{\rho_{\rm{DPXA}}(s)}\leq1$ for DPXA. The DPXA coefficient
indicates the intrinsic cross-correlations between two non-stationary
series.

\begin{figure}[t]
  \centering
  \includegraphics[width=0.95\linewidth]{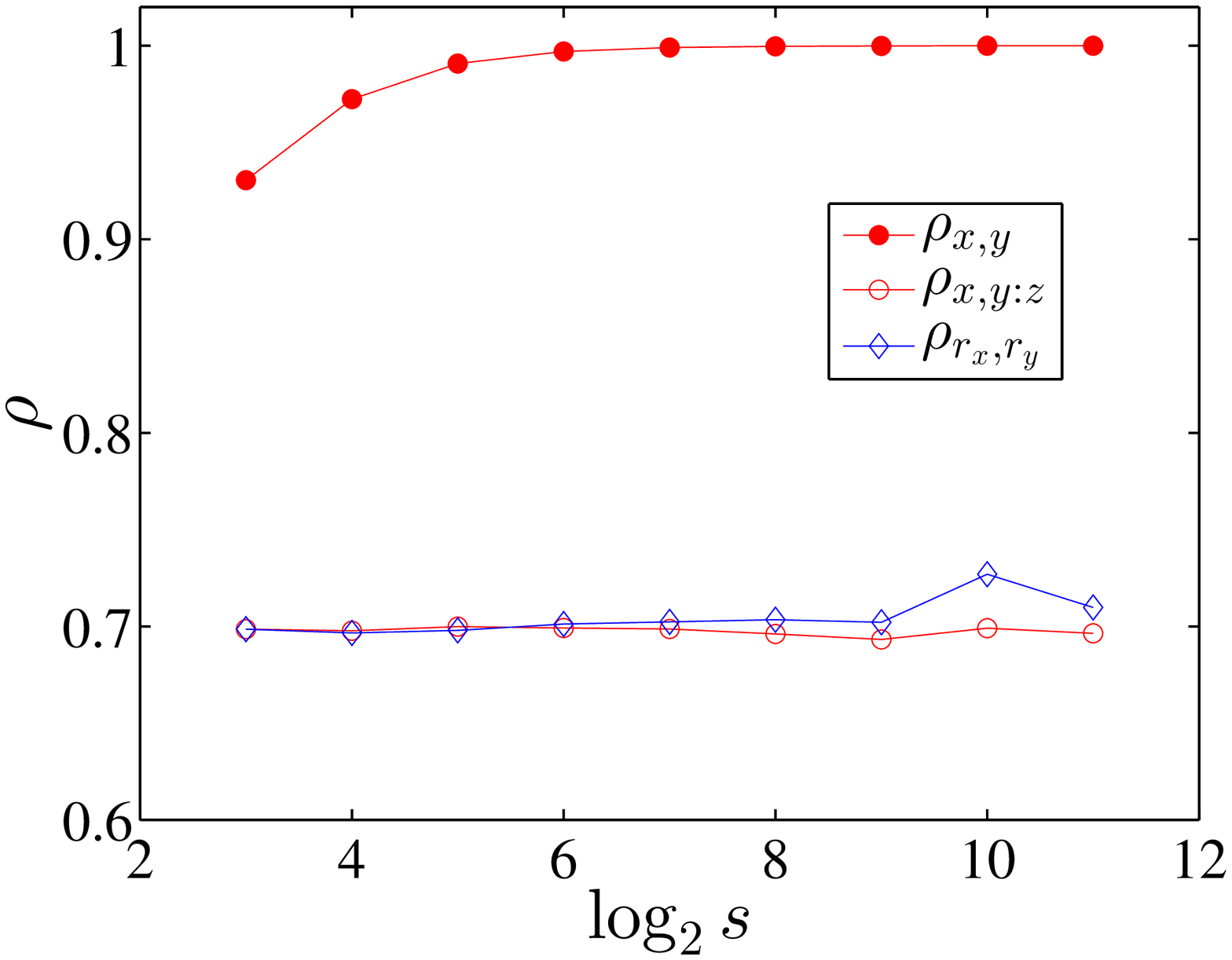}
  \includegraphics[width=0.95\linewidth]{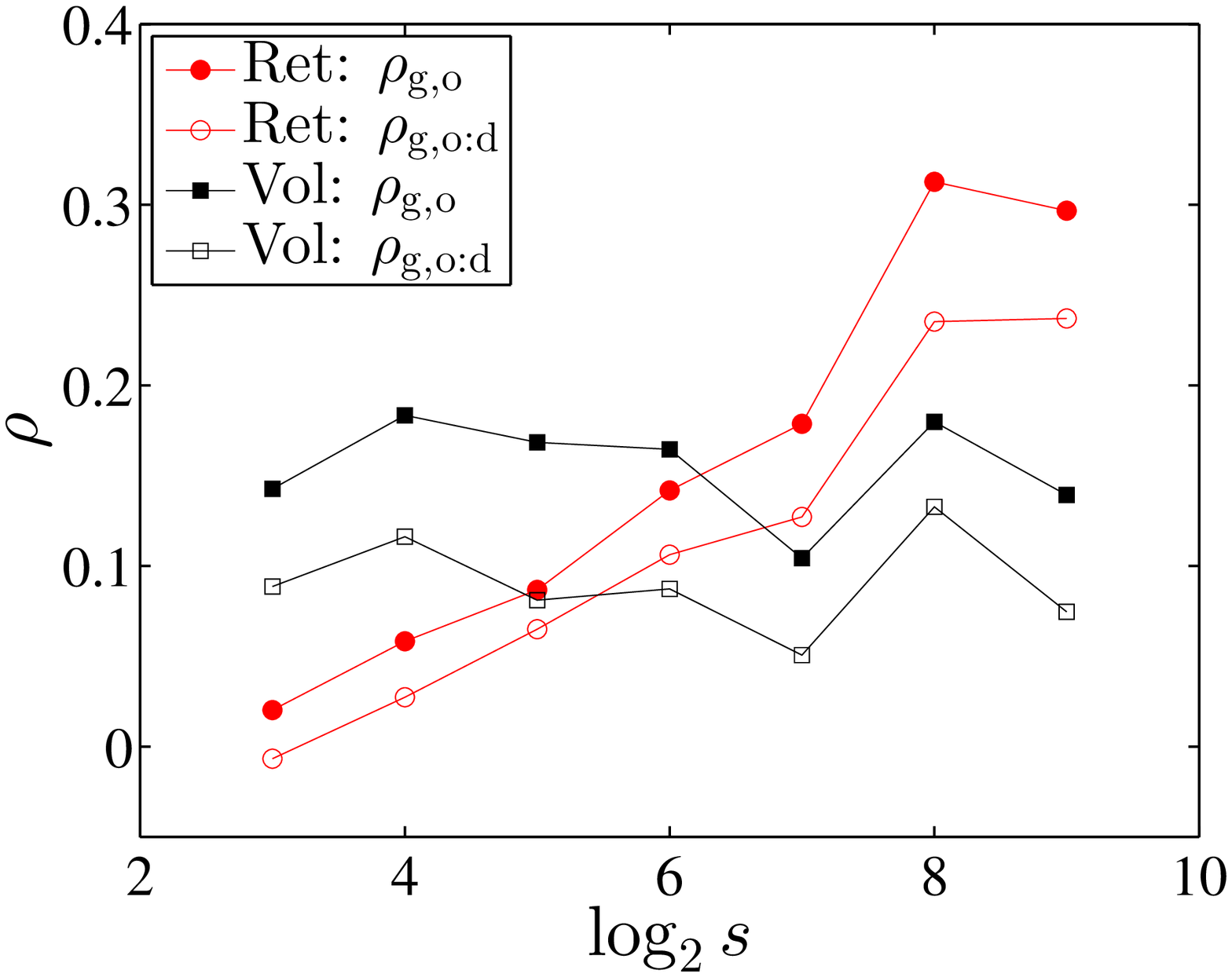}
  \caption{\label{Fig:DPXA:rho} (color online.) Detrended partial
    cross-correlation coefficients. (a) Performance of different methods
    by comparing three cross-correlation coefficients $\rho_{x,y}$,
    $\rho_{r_x,r_y}$ and $\rho_{x,y:z}$ of the mathematical model in
    Eq.~(\ref{Eq:DPXA:Model}). (b) Estimation and comparison of the
    cross-correlation levels between the two return time series
    ($\circ$) and two volatility time series ($\square$) of crude oil
    and gold when including and excluding the influence of the USD
    index.}
\end{figure}

We use the mathematical model in Eq.~(\ref{Eq:DPXA:Model}) with the
coefficients $\beta_{x,0}=\beta_{y,0}=2$ and $\beta_{x,1}=\beta_{y,1}=3$
to demonstrate how the DPXA coefficient outperforms the DCCA
coefficient. The two components $r_x$ and $r_y$ of the BFBM have very
small Hurst indices $H_{r_x}=0.1=H_{r_y}= 0.1$ and their correlation
coefficient is $\rho=0.7$, and the driving FBM force $z$ has a large
Hurst index $H_z= 0.95$. Figure~\ref{Fig:DPXA:rho}(a) shows the
resulting cross-correlation coefficients at different scales. The DCCA
coefficients $\rho_{r_x,r_y}$ between the generated $r_x$ and $r_y$ time
series overestimate the true value $\rho=0.7$. Because the influence of
$z$ on $r_x$ and $r_y$ is very strong, the behaviors of $x$ and $y$ are
dominated by $z$, and the cross-correlation coefficient $\rho_{x,y}(s)$
is close to 1 when $s$ is small and approaches 1 when $s$ us large. In
contrast, the DPXA coefficients $\rho_{x,y:z}$ are in good agreement
with the true value $\rho=0.7$. Note that the DPXA method better
estimates $r_x$ and $r_y$ than the DCCA method, since the
$\rho_{r_x,r_y}$ curve deviates more from the horizontal line $\rho=0.7$
than the $\rho_{x,y:z}$ curve, especially at large scales.

To illustrate the method with an example from finance, we use it to
estimate the intrinsic cross-correlation levels between the futures
returns and the volatilities of crude oil and gold. It is
well-documented that the returns of crude oil and gold futures are
correlated \cite{Zhang-Wei-2010-RP}, and that both commodities are
influenced by the USD index \cite{Wang-Chueh-2013-EM}. The data samples
contain the daily closing prices of gold, crude oil, and the USD index
from 4 October 1985 to 31 October 2012. Figure \ref{Fig:DPXA:rho}(b)
shows that both the DCCA and DPXA coefficients of returns exhibit an
increasing trend with respect to the scale $s$, and that the two types
of coefficient for the volatilities do not exhibit any evident
trend. For both financial variables, Fig.~\ref{Fig:DPXA:rho}(b) shows
that
\begin{equation}
  \rho_{\mathrm{g,o:d}}(s)<\rho_{\mathrm{g,o}}(s)
  \label{Eq:rho:DCCA:DPXA}
\end{equation}
for different scales. Although this is similar to the result between
ordinary partial correlations and cross-correlations
\cite{Kenett-Huang-Vodenska-Havlin-Stanley-2015-QF}, the DPXA
coefficients contain more information than the ordinary partial
correlations since the former indicate the partial correlations at
multiple scales.

\begin{figure*}[!ht]
  \centering
  \includegraphics[width=0.95\linewidth]{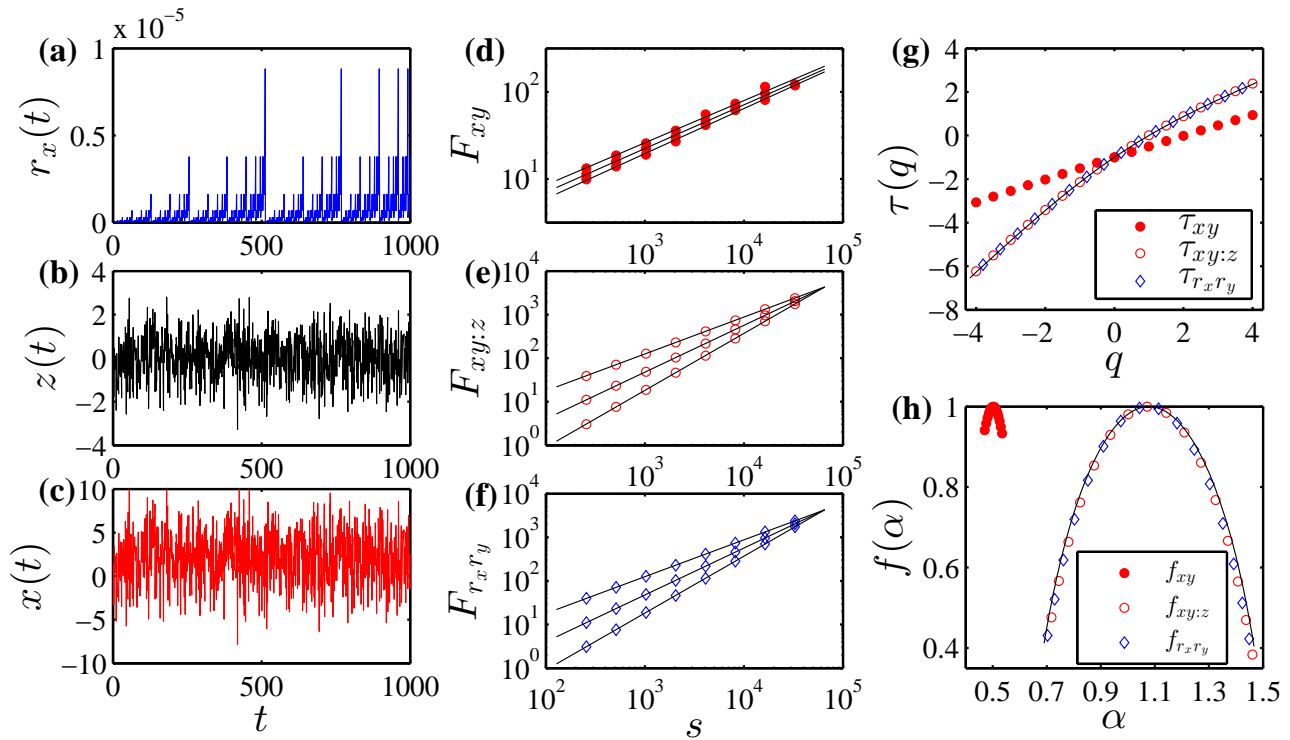}
  \caption{\label{Fig:MFPX:pmodel} (color online.) Multifractal
    detrended partial cross-correlation analysis of two binomial
    measures contaminated by Gaussian noise with very low
    signal-to-noise ratio. (a-c) The segments of the binomial signal
    $r_x(t)$ with $p_x=0.3$, the Gaussian noise $z(t)$, and the
    ``observed'' signal $x(t)$. (d-f) Power-law dependence of the
    fluctuations $F_{xy}(q,s)$, $F_{xy:z}(q,s)$, $F_{r_xr_y}(q,s)$ on
    the scale $s$ for $q=-4$, 0 and 4 (top down). The values of
    $F_{xy:z}$ and $F_{r_xr_y}$ have been multiplied by $10^5$. (g)
    Multifractal mass exponents $\tau_{xy}(q)$, $\tau_{xy:z}(q)$ and
    $\tau_{r_xr_y}(q)$, with the theoretical curve ${\cal{T}}_{r_xr_y}$
    shown as a continuous line. (d) Multifractal spectra
    $f_{xy}(\alpha)$, $f_{xy:z}(\alpha)$, $f_{r_xr_y}(\alpha)$ of the
    singularity strength $\alpha$. The continuous curve is the
    theoretical spectrum ${\cal{F}}_{r_xr_y}(\alpha)$.}
\end{figure*}

\section{Multifractal detrended partial cross-correlation analysis}

An extension of the DPXA for multifractal time series, notated MF-DPXA,
can be easily implemented. When MF-DPXA is implemented with DFA or DMA,
we notate it MF-PX-DFA or MF-PX-DMA. The $q$th order detrended partial
cross-correlation is calculated
\begin{equation}
 F_{xy:{\bm{z}}}(q,s) = \left[\frac{1}{m-1}\sum_{v=1}^m |F_{v}^2(s)|^{q/2}\right]^{1/q}
\end{equation}
when $q\neq0$, and
\begin{equation}
 F_{xy:{\bm{z}}}(0,s) = \exp\left[\frac{1}{m}\sum_{v=1}^m \ln |F_{v}(s)|\right]~.
\end{equation}
We then expect the scaling relation
\begin{equation}
 F_{xy:{\bm{z}}}(q,s) \sim s^{h_{xy:{\bm{z}}}(q)}~.
 \label{Eq:Fxy:q:s}
\end{equation}
According to the standard multifractal formalism, the multifractal mass
exponent $\tau(q)$ can be used to characterize the multifractal nature,
i.e.,
\begin{equation}
\tau_{xy:{\bm{z}}}(q)=qh_{xy:{\bm{z}}}(q)-D_f,
\label{Eq:MFDCCA:tau}
\end{equation}
where $D_f$ is the fractal dimension of the geometric support of the
multifractal measure
\cite{Kantelhardt-Zschiegner-KoscielnyBunde-Havlin-Bunde-Stanley-2002-PA}. We
use $D_f=1$ for our time series analysis. If the mass exponent $\tau(q)$
is a nonlinear function of $q$, the signal is multifractal. We use the
Legendre transform to obtain the singularity strength function
$\alpha(q)$ and the multifractal spectrum $f(\alpha)$
\cite{Halsey-Jensen-Kadanoff-Procaccia-Shraiman-1986-PRA}
\begin{equation}
    \left\{
    \begin{array}{ll}
        \alpha_{xy:{\bm{z}}}(q)={\rm{d}}\tau_{xy:{\bm{z}}}(q)/{\rm{d}}q\\
        f_{xy:{\bm{z}}}(q)=q{\alpha_{xy:{\bm{z}}}}-{\tau_{xy:{\bm{z}}}}(q)
    \end{array}
    \right..
\label{Eq:MFDCCA:f:alpha}
\end{equation}

To test the performance of MF-DPXA, we construct two binomial measures
$\{r_x(t): t = 1, 2, \cdots, 2^k\}$ and $\{r_y(t): t = 1, 2, \cdots,
2^k\}$ from the $p$-model with known analytic multifractal properties
\cite{Meneveau-Sreenivasan-1987-PRL}, and contaminate them with Gaussian
noise. We generate the binomial measure iteratively
\cite{Jiang-Zhou-2011-PRE} by using the multiplicative factors $p_x=0.3$ for
$r_x$ and $p_y=0.4$ for $r_y$. The contaminated signals are $x=2+3
z+r_x$ and $y=2+3 z+r_y$.  Figures~\ref{Fig:MFPX:pmodel}(a)--(c) show
that the signal-to-noise ratio is of order
$O(10^{-6})$. Figures~\ref{Fig:MFPX:pmodel}(d)--(f) show a power-law
dependence between the fluctuation functions and the scale, in which it
is hard to distinguish the three curves of
$F_{xy}$. Figure~\ref{Fig:MFPX:pmodel}(g) shows that for $x(t)$ and
$y(t)$, the $\tau_{xy}(q)$ function an approximate straight line and
that the corresponding $f_{xy}(\alpha)$ spectrum is very narrow and
concentrated around $\alpha=0.5$. These observations are trivial because
$x(t)$ and $y(t)$ are Gaussian noise with the Hurst indices
$H_x=H_y=0.5$, and the multifractal detrended cross-correlation analysis
\cite{Zhou-2008-PRE} fails to uncover any multifractality. On the
contrary, we find that
$\tau_{xy:z}(q)\approx\tau_{r_xr_y}(q)\approx{\cal{T}}_{r_xr_y}(q)$ and
$f_{xy:z}(\alpha)\approx
f_{r_xr_y}(\alpha)\approx{\cal{F}}_{r_xr_y}(\alpha)$. Thus the MF-DPXA
method successfully reveals the intrinsic multifractal nature between
$r_x(t)$ and $r_y(t)$ hidden in $x(t)$ and $y(t)$.

\section{Summary}

In summary, we have studied the performances of DPXA exponents, DPXA coefficients, and MF-DPXA using bivariate fractional Brownian motions contaminated by a fractional Brownian motion and multifractal binomial measures contaminated by white noise. These mathematical models are appropriate here because their analytical expressions are known. We have demonstrated that the DPXA methods are capable of extracting the intrinsic cross-correlations between two time series when they are influenced by common factors, while the DCCA methods fail.

The methods discussed are intended for multivariate time series analysis, but they can also be generalized to higher dimensions \cite{Gu-Zhou-2006-PRE,Carbone-2007-PRE,Gu-Zhou-2010-PRE,Jiang-Zhou-2011-PRE}. We can also use lagged cross-correlations in these methods \cite{Podobnik-Wang-Horvatic-Grosse-Stanley-2010-EPL,Shen-2015-PLA}. Although comparing the performances of different methods is always important \cite{Shao-Gu-Jiang-Zhou-Sornette-2012-SR}, different variants of a method can produce different outcomes when applied to different systems. For instance, one variant that outperforms other variants under the setting of certain stochastic processes is not necessary the best performing method for other systems \cite{Gu-Zhou-2010-PRE}. We argue that there are still a lot of open questions for the big family of DFA, DMA, DCCA and DPXA methods.

\begin{acknowledgments}

This work was partially supported by the National Natural Science
Foundation of China under grant no. 11375064, Fundamental Research Funds
for the Central Universities, and Shanghai Financial and Securities
Professional Committee.

\end{acknowledgments}


%

\end{document}